\documentclass[12pt,preprint]{aastex}
\usepackage[dvips]{epsfig}
\usepackage{paralist}
\usepackage[latin1]{inputenc}
\slugcomment{Draft version}
\shorttitle{Maximal ratio}
\shortauthors{Arreaga-Garcia}

\begin{document}
\title{The extreme initial kinetic energy allowed by a collapsing turbulent core} 

\author{Guillermo Arreaga-Garc\'{\i}a\altaffilmark{1}}
\affil{Departamento de Investigaci\'on en F\'{\i}sica, Universidad de Sonora, \\
Apdo. Postal 14740, C.P. 83000, Hermosillo, Sonora, Mexico.}

\begin{abstract}
We present high-resolution hydrodynamical simulations aimed
at following the gravitational collapse of a gas core, in which a turbulent spectrum 
of velocity is implemented only initially. We determine the maximal value of the 
ratio of kinetic energy to gravitational energy, denoted here by 
$\left(\frac{E_{\rm kin} }{E_{\rm grav}}\right)_{\rm max}$, so that the core (i) will collapse 
around one free-fall time of time evolution or (ii) will expand unboundedly, because it has a value 
of $\frac{E_{\rm kin}}{E_{\rm grav}}$ larger than 
$\left( \frac{E_{\rm kin}}{E_{\rm grav}}\right)_{\rm max}$. We consider core models with a uniform or
centrally condensed density profile and with velocity spectra composed 
of a linear combination of one-half divergence-free 
turbulence type and the other half of a curl-free turbulence type.
We show that the outcome of the core collapse are
protostars forming either (i) a multiple system obtained from the fragmentation of
filaments and (ii) a single primary system within a long filament. In addition, some properties
of these protostars are also determined and compared with those obtained elsewhere.
\end{abstract}
\section{Introduction}
\label{sec:intro}

The formation of stars begins in the interstellar medium when a
cloud of molecular hydrogen becomes gravitationally unstable, so that
it collapses simultaneously in many regions of the cloud. Thus, many
gas condensations are formed in the cloud, which are usually referred to as
pre-stellar gas cores, see \cite{bergin}.

Much effort has been expended to understand the last part of
this process, namely, the collapse of the cores. For instance, \cite{whithworth}
considered the dynamical evolution of a collapsing pre-stellar core using an
analytical model with the assumptions that the gas has (i) negligible
pressure and rotation and (ii) a Plummer-like radial density profile similar to the one
introduced by \cite{plu}. \cite{whithworth} concluded that this
Plummer-like model captures quite well some of the observed properties of
the dense core L1544.

The core L1544 was also considered by \cite{goodwin2004a} and
\cite{goodwin2004b}, who modeled it numerically, so that the
collapse of the core was triggered by implementing a divergence-free
turbulent spectrum. In addition, \cite{goodwin2006} studied the
influence of different levels of divergence-free turbulence on the
fragmentation and multiplicity of these cores. These papers sought for how low can
the levels of turbulence be to still favour core fragmentation. Moreover,
\cite{attwood} obtained a better modeling of the thermodynamics of
the core collapse by introducing an energy equation, whose results
were compared with the ones obtained from simulations using the
barotropic equation of state, such as those of \cite{goodwin2004a},
\cite{goodwin2004b}, and \cite{goodwin2006}.

\cite{walch} considered a mixed turbulent velocity
spectrum (with a ratio of divergence-free type to curl-free type of 2:1) such
that a cubic mesh of 128$^3$ grid elements was
populated with Fourier modes, in order to calculate the collapse of a core of radius
$R_0$ under the influence of modes with wavelength
$\lambda_{\rm max}$ within the range $R_0$/2, $R_0$, 2 $R_0$
and 4 $R_0$. It must be mentioned that \cite{walch} observed core
fragmentation only for the models with $R_0/2 \leq \,
\lambda_{\rm max} \, \leq 2 R_0$.

In this paper, we study the gravitational collapse of a pre-stellar core which can have a density profile
either uniform or Plummer-like, such as the one proposed by \cite{whithworth} and also calculated
by \cite{goodwin2004a}, \cite{goodwin2004b}, and \cite{goodwin2006}. Our models include a
linear combination of two extreme types of turbulent velocity spectra, so that $\vec{v}(\vec{r}) =
\frac{1}{2} \, \vec{v}_{\rm DF}(\vec{r}) + \frac{1}{2} \, \vec{v}_{\rm CF}(\vec{r})$ where
$\vec{v}_{\rm DF}$ is a divergence-free turbulent spectrum (also called solenoidal) and
$\vec{v}_{\rm CF}$ is a curl-free turbulent spectrum (also called compressive).

It must be mentioned that most of the simulations done worldwide
about the collapse of turbulent cores have considered only
solenoidal turbulence. In the context of gas clouds, \cite{fede}
presented a very detailed statistical comparison of the properties
not only of the two types of turbulence considered separately, but
also of a mixed type of turbulence that includes any desired ratio
of both solenoidal and compressive types. In the context of the
collapse of turbulent cores, \cite{giri2011} have studied the influence of four 
different density profiles and considered many different initial turbulent 
velocity fields: solenoidal, compressive and mixed turbulent fields in order 
to study star and star cluster formation by means of fragmentation of turbulent cores.

In addition, \cite{RMAA2017} has considered
separately the collapse of cores having each type of turbulence.
\cite{lomax} followed the evolution of pre-stellar cores, in
which the fraction of turbulent energy varied in five different
combinations of the velocity field. The velocity field proposed in
this paper is proportional to model 2 of \cite{lomax}, so that the
ratio of the coefficients in front of the turbulent velocity spectra
is 1.

The Fourier modes considered in this paper take values on a cubic mesh of 128$^3$
grid elements, and the range of wavelength $\lambda_{\rm max}$ considered here is larger
than that used by \cite{walch}, so that it takes values from 1, 4 and 10 times the core
radius $R_0$, as was done by \cite{RMAA2017}, where the effects on the core collapse 
of changes in the number and size of turbulent modes of velocity were studied.

We emphasize that the mentioned papers of \cite{goodwin2004a},
\cite{goodwin2004b}, and \cite{goodwin2006} as well as those of
\cite{bate2002a}, \cite{bate2002b} and \cite{bate2003}, who
presented collapse simulations in the context of gas clouds and
using only solenoidal turbulence, all suggested that
turbulent fragmentation can be a natural and efficient mechanism for
explaining the formation of binary systems. In addition,
\cite{padoan2002} demonstrated that the observed shape of the
stellar initial mass function can be correctly obtained from
collapse simulations of supersonically turbulent molecular clouds, see the review 
of \cite{hopkins}. 

In this paper, we focus only on decaying turbulence, as has been done
in most of the simulations of turbulent cores, see \cite{goodwin2004a},
\cite{goodwin2004b}, and \cite{goodwin2006}. \cite{fede} considered the
evolution of molecular clouds under the influence of driven turbulence, so
that it is maintained for all the time of evolution, although their simulations did
not include self-gravity. We emphasize that, as far as we know, a collapse
calculation of an isolated core under driven turbulence is still missing.

We emphasize that simulations of the collapse of turbulent gas is a very active field of
research, in which the more advanced physical ideas and computational techniques have been 
tested; see for instance the review of \cite{padoan2014}, who reported
recent advances focusing on the connection of the physics of turbulence
with the star formation rate in molecular clouds. In addition, \cite{giri2011} 
included the sink particle technique in modeling the collapse of turbulent cores; 
\cite{fede2012} considered numerical models of forced, 
supersonic, self-gravitating, magneto-hydrodynamical turbulence 
to study the fragmentation of clouds; \cite{fede2015} considered the effects of 
turbulence, magnetic fields and feedback on the star formation rate. 
The simulations by \cite{fede2012} and \cite{fede2015} included driven turbulence 
and modeled clouds in which cores form automatically instead of starting from an idealized 
initial condition with an isolated spherical core.

For all the simulations of this paper, the ratio of the thermal energy to the
gravitational energy is fixed at 0.24. The models have
been calibrated so that the ratio of the kinetic energy to the gravitational
energy takes its maximal value consistent with a collapsing core. We also
calculate some properties of the resulting protostars, such as the mass $M_f$ 
and the ratios
$\left( \frac{E_{\rm ther}}{E_{\rm grav}}\right)_f$ 
and $\left( \frac{E_{\rm kin}}{E_{\rm grav}}\right)_f$ 
and compare our results with those obtained from the collapse of rotating cores,
see ~\cite{miApJ},\cite{miAA}, \cite{RMAA2016} and \cite{miAAS}.

The time evolution of our models is achieved by using the publicly available
code Gadget2, which is based on the Smoothed Particle Hydrodynamics (SPH) technique for
solving the hydrodynamic equations coupled to self-gravity. When gravity has produced a
substantial contraction of the core, the gas begins to
heat. To take this increase of temperature into account, we
use a barotropic equation of state, which was first proposed by
\cite{boss2000}.

It must be mentioned that some physical properties of the core
considered in this paper are the same as  those considered by \cite{goodwin2004a}
and \cite{goodwin2004b}, namely, the radius, the mass, the Plummer-like density
profile, and the equation of state. The elements to be emphasized in our simulations
that differ from those of \cite{goodwin2004a} and \cite{goodwin2004b} are the
following: (i) the addition of a curl-free turbulence term, since
they only considered a divergence-free term; (ii) the search for the maximal value of
$\frac{E_{\rm kin}}{E_{\rm grav}}$ that a collapsing core allowed, whereas they 
searched for the minimal values
of turbulent energy that allowed core fragmentation; (iii) a significant increase of the
number of simulation particles, as they used in general 25,000 and at most
100,000 SPH particles, whereas all our simulations have a little more than
10,000,000 SPH particles.

\section{The physical system and the computational method}
\label{subsec:met}

\subsection{The core}
\label{subsec:core}

The core considered in this paper has a radius of $R_0= 0.242$ pc $\equiv$ 49933.86 AU
and a mass of $M_0=5.41 \, M_{\odot}$. Thus, the average density and the corresponding free
fall time of this core are $\rho_0$=6.16 $\times 10^{-21}$ g
cm$^{-3}$ and $t_{ff} \approx $2.67 $\times 10^{13} \,$ s or
0.84 Myr, respectively. These values of $R_0$ and $M_0$
have been taken from \cite{goodwin2004a} in order to make comparisons.

In this paper, we will consider two kinds of density profiles: uniform and
centrally condensed, according to the following functions:

\begin{equation}
\begin{array}{l}
\rho(r)=\rho_0 \; \vspace{0.1 cm}\\
\rho(r)=\rho_c \,  \left( \frac{R_c}{\sqrt{r^2+R_c^2}} \right)^\eta \\
\end{array}
\label{distdens}
\end{equation}
\noindent The second formula
was first introduced by \cite{plu} and later on studied
by \cite{whithworth} in the context of the theory of star formation. This function
includes three free parameters: $\rho_c$, which establishes the central density
value; a critical radius $R_c$ that sets up the
end of the approximately constant part of the radial density curve
for the innermost matter ($0 < r \leq R_c$); and an exponent $\eta$
that fixes the density fall rate for a large radius ($r \gg R_c$).

In this paper, $\eta$ is fixed at the value of 4, as was constrained by the observational
lifetimes of cores; see \cite{whithworth}. Meanwhile, $R_c$ will take
values proportional to $R_0$, so that we will have two kinds of
centrally condensed models: those with $R_c=R_0/2.5$, which will be referred to as C models;
and those having $R_c=R_0/10$, which will be referred to as R models; see
Section~\ref{subs:mod} and Table~\ref{tab:mod}.

In order to have a set of particles that will reproduce
the desired density profiles, we proceed as follows.
We make a partition of the simulation volume into small cubic elements, each with a
volume $\Delta x\, \Delta y\, \Delta z $; at the centre of
each cubic element we place a particle. Next, we displace each particle a distance of
the order $\Delta x/4.0$ in a random spatial direction within each cubic
element.

The simulation particles get a mass according to
the density profiles shown in Eq. \ref{distdens}, so that particle $i$
has mass $m_i= \rho(r_i) \times \Delta x\, \Delta y\, \Delta z$, for
$i=1,...,N_p$, where $N_p$ is the total number of particles, which was set
to be 10,034,074 in order to fulfill the resolution requirements; see Section~\ref{subs:resol}.

The density perturbation was achieved here by means of a mass perturbation, as was remarkably done
by \cite{gadget2}.

\begin{figure}
\begin{center}
\includegraphics[width=4.0 in]{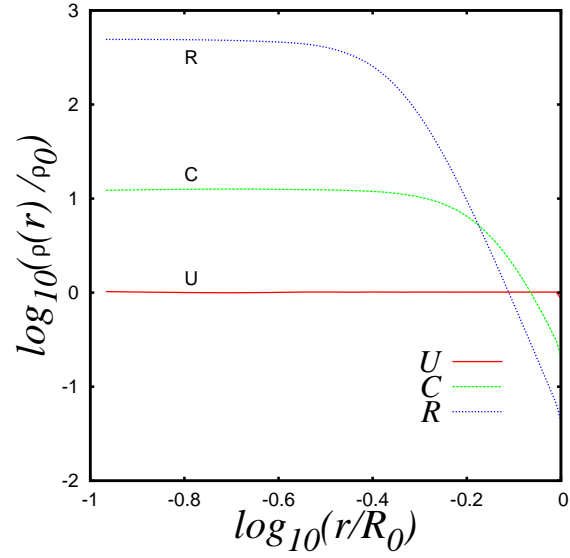}
\caption{\label{fPerfil} Radial density profile for the initial distribution of particles.}
\end{center}
\end{figure}

It should be noted that all the models to be presented in Section \ref{subs:mod} have a total core mass
of $M_0= 5.41 \, M_{\odot}$. Therefore, the particle masses $m_i$ are affected by a
multiplicative constant, whose value obviously depends on the model in consideration. The value of the central
density, $\rho_c$, is also affected. Thus, in Figure~\ref{fPerfil} we present the initial
density profiles for the uniform and the two kinds of centrally condensed models
considered here, as they were numerically measured from the initial snapshot.

\subsection{The velocity of the particles}
\label{subsec:velfin}

The velocity vector $\vec{v}(\vec{r})$ to be given to each SPH particle located
at position $\vec{r}$ will be formed by the following combination of the two types
of turbulent spectra, so that

\begin{equation}
\vec{v}(\vec{r}) = \frac{1}{2} \, \vec{v}_{\rm DF}(\vec{r}) + \frac{1}{2} \, \vec{v}_{\rm CF}(\vec{r})
\label{velfin}
\end{equation}
\noindent The level of turbulence is adjusted by introducing a multiplicative constant in front of
the right hand side of Eq. \ref{velfin}. It should be noted that \cite{RMAA2017} examined
the effects on the collapse of cores due to variation of the
number and size of the Fourier modes, although each turbulence
type was allowed to act upon the core separately.

To generate the turbulent velocity spectrum, we set a second
mesh, with a side length denoted here by $L_0$, which is proportional to the core
radius $R_0$, so that

\begin{equation}
L_0=C_R \times R_0
\label{defCR}
\end{equation}
\noindent where $C_R$ is a constant, the value of which
will also determine the collapse model under consideration, see Section~\ref{subs:mod} and
Table~\ref{tab:mod}.

It must be noted that each term of Eq. \ref{velfin} was calculated in Section
2.4 of \cite{RMAA2017}, so that we do not repeat it here and show only the final expressions.

The components of the first term of Eq. \ref{velfin} are given by

\begin{equation}
\begin{array}{l}

\vec{v}_{\rm DF} \, \approx \Sigma_{i_x,i_y,i_z} \left| \vec{K}\right|^{ \frac{-n-2}{2} }
\times

\left\{

\begin{array}{l}

\left[ K_z \, C_{K_y} \sin \left( \vec{K}\cdot \vec{r} + \Phi_{K_y}\right) -
K_y \, C_{K_z} \sin \left( \vec{K}\cdot \vec{r} + \Phi_{K_z}\right)\right] \; \mbox{for}\, v_x\\

\left[ - K_x \, C_{K_z} \sin \left( \vec{K}\cdot \vec{r} + \Phi_{K_z}\right) +
K_z \, C_{K_x} \sin \left( \vec{K}\cdot \vec{r} + \Phi_{K_x}\right)\right] \; \mbox{for} \, v_y\\

\left[ -K_x \, C_{K_y} \sin \left( \vec{K}\cdot \vec{r} + \Phi_{K_y}\right) +
K_y \, C_{K_x} \sin \left( \vec{K}\cdot \vec{r} + \Phi_{K_x}\right)\right] \; \mbox{for} \, v_z\\
\end{array}

\right.
\end{array}
\label{velturb}
\end{equation}
\noindent see \cite{dobbs}. For the second term of Eq. \ref{velfin} we have

\begin{equation}
\vec{v}_{\rm CF}(\vec{r})  \approx \Sigma_{i_x,i_y,i_z} \left| \vec{K}\right|^{\frac{-n-2}{2}} \;
\vec{K} \sin \left( \vec{K}\cdot \vec{r} + \Phi_K \right)
\label{velPhi}
\end{equation}
\noindent where the spectral index $n$ has been fixed in all our
simulations to $n=-1$.

In the literature on turbulence, a parameter $\xi \in$ [0,1] is introduced to determine
the relative contribution of the divergence-free and curl-free turbulent modes to the velocity
field, see \citet{fede} and \citet{bruntyfede}. According to equation 61 of \citet{bruntyfede}, the value
$\xi=$1/2, as the one chosen in this paper, implies that the ratio of
curl-free power to total power of the velocity field is given by 1/3. \citet{lomax} used
the parameter $\delta_{\rm sol}$ that characterize the ratio of turbulent energy in diverge-free
turbulent modes to the total turbulent energy, so that their model 2 defined by a velocity field
given by $\vec{v}_{\rm DF} + \vec{v}_{\rm CF}$ has $\delta_{\rm sol}=$2/3, which
is called a thermal mixture of turbulent modes. As we mentioned earlier, the velocity combination of this paper
shown in Eq. \ref{velfin}, can be compared to model 2 of \citet{lomax}.


\subsection{Initial energies}
\label{subs:energies}

The relevant energies are  calculated using all the SPH particles as follows:

\begin{equation}
\begin{array}{l}
E_{\rm ther}=\frac{3}{2}\sum_{i} \, m_{i}\frac{P_{i}}{\rho _{i}}\\
E_{\rm kin}=\frac{1}{2}\sum_{i} \, m_{i} v_i^{2},\\
E_{\rm grav}=\frac{1}{2}\sum_{i} \, m_{i}\Phi_{i}
\label{energiespart}
\end{array}
\end{equation}
\noindent where $P_i$ and $\Phi _{i}$ are the values
of the pressure and gravitational potential at the location of particle
$i$, with velocity given by $v_i$ and mass $m_i$; the summations
include all the simulation particles.

In this paper, the value of the speed of sound $c_0$ is chosen in each model so
that the simulations have

\begin{equation}
\frac{E_{\rm ther}}{\left|E_{\rm grav}\right|} \equiv 0.24
\label{valpha}
\end{equation}
\noindent It should be noted that there are three different values of
the speed of sound, see column 6 of Table~\ref{tab:mod}. Thus, the corresponding temperatures associated
with the core are $T \approx 3,4,7 \, $K, respectively. It should be mentioned
that \cite{goodwin2004a} used a value of $\frac{E_{\rm ther}}{\left|E_{\rm grav}\right|} \equiv$ 0.45.

The level of turbulence is chosen to get the maximal value
of $\frac{E_{\rm kin} }{E_{\rm grav}}$ in each model (see Section\ref{subs:mod} and Table~\ref{tab:mod}), so that this
is the maximum value for core collapse. In Section \ref{subs:mod}, we will explain how the
values of $\left(\frac{E_{\rm kin}}{E_{\rm grav}}\right)_{\rm max}$ are obtained.

It must be mentioned that the virial theorem for a closed system in thermodynamic equilibrium
can be expressed by the following relation:

\begin{equation}
\frac{E_{\rm ther} }{E_{\rm grav}} + \frac{E_{\rm kin} }{E_{\rm grav}}=\frac{1}{2}\;,
\label{abvirial}
\end{equation}

It should be mentioned that these energy ratios play a significant role in the determination of the stability of a
gaseous system against gravitational perturbations. According to the virial theorem, if a gaseous system
has $\frac{E_{\rm ther} }{E_{\rm grav}} + \frac{E_{\rm kin} }{E_{\rm grav}} > 1/2$, then it will expand; in 
the other case, if $\frac{E_{\rm ther} }{E_{\rm grav}} + \frac{E_{\rm kin} }{E_{\rm grav}}<1/2$, then
the system will collapse. \cite{miyama}, \cite{hachisu1} and \cite{hachisu2} obtained a more precise
criterion of the type $\frac{E_{\rm ther} }{E_{\rm grav}} \times \frac{E_{\rm kin} }{E_{\rm grav}}<  0.2 $ to predict the
collapse and fragmentation of a rotating isothermal core.
\subsection{Evolution Code}
\label{subs:code}

The gravitational collapse of our models has been followed by using
the fully parallelized particle-based code Gadget2;
see~\cite{gadget2} and also \cite{serial}. Gadget2 is based on the
tree-PM method for computing the gravitational forces and on the
standard smoothed particle hydrodynamics (SPH)  method for solving
the Euler equations of hydrodynamics. Gadget2 implements a
Monaghan--Balsara form for the artificial viscosity;
see~\cite{mona1983} and \cite{balsara1995}. The strength of the
viscosity is regulated by the parameter $\alpha_{\nu} = 0.75$ and
$\beta_{\nu}=\frac{1}{2}\, \times \alpha_v$; see Eqs. 11 and 14
in~\cite{gadget2}. In our simulations we fixed the Courant factor
to be $0.1$.
\subsection{Resolution and equation of state}
\label{subs:resol}

The reliability of a program in calculating the collapse is determined by the resolution
needed, which can be expressed in terms of the Jeans wavelength $\lambda_J$:

\begin{equation}
\lambda_J=\sqrt{ \frac{\pi \, c^2}{G\, \rho}} \; , \label{ljeans}
\end{equation}
\noindent where $c$ is the instantaneous speed of sound and $\rho$ is the
local density; or to obtain a more useful form for a particle based code, the Jeans
wavelength $\lambda_J$ can be transformed into a Jeans mass, given by

\begin{equation}
M_J \equiv \frac{4}{3}\pi \; \rho \left(\frac{ \lambda_J}{2}
\right)^3 = \frac{ \pi^\frac{5}{2} }{6} \frac{c^3}{ \sqrt{G^3 \,
\rho} } \;. \label{mjeans}
\end{equation}
\noindent where the values of the density and speed of sound must be updated according to
the following equation of state:

\begin{equation}
p= c_0^2 \, \rho \left[ 1 + \left(
\frac{\rho}{\rho_{crit}}\right)^{\gamma -1 } \, \right] ,
\label{beos}
\end{equation}
\noindent which was proposed by \cite{boss2000}, where $\gamma\,
\equiv 5/3$ and for the critical density we assume the value
$\rho_{crit}=5.0 \times 10^{-14} \, $ g $\,$ cm$^{-3}$.

Thus, the smallest mass particle that a SPH calculation must resolve in
order to be reliable is given by $m_r \approx M_J / (2 N_{\rm neigh})$, where
$N_{\rm neigh}$ is the number of neighbouring particles included in the
SPH kernel; see \cite{truelove} and \cite{bateburkert97}. Hence, a simulation
satisfying all the resolution requirements must satisfy $\frac{m_p}{m_r}<1$.

For the turbulent core under consideration, the number of
particles is $N_p$ = 10,034,074, so that the average particle mass is given by
m$_p$=5.3 $\times \, 10^{-7} \, $ M$_{\odot}$. In addition, the number of neighboring
particles is $N_{\rm neigh}=40$.

Let us first consider the U models, with the smallest value of the
speed of sound, as given in Table~\ref{tab:mod}, and assume that the peak
density of these simulations to be used in
this resolution calculation is $\rho=2.5 \times 10^{-12}$ g cm$^{-3}$,
then $M_j \approx 4.8 \times 10^{-5} \,$ M$_{\odot}$ and
m$_r \approx 6 \times 10^{-7}\,$ M$_{\odot}$. In this case, the ratio of masses is given by $m_p/m_r=$0.885
and then the desired resolution is achieved in these simulations up to this level of
peak density. For the C models and assuming a peak density of $\rho=1.0 \times 10^{-11}$ g cm$^{-3}$,
we have $M_j \approx 4 \times 10^{-5} \,$ M$_{\odot}$ and
m$_r \approx 5 \times 10^{-7}\,$ M$_{\odot}$, so that $m_p/m_r=$1.0. Now consider the R models, which
have the largest value of the sound speed,
as given in Table~\ref{tab:mod}, and
again $\rho=2.5 \times 10^{-11}$ g cm$^{-3}$, then
$M_j \approx 8.5 \times 10^{-5} \,$ M$_{\odot}$ and m$_r \approx 1.6
\times 10^{-6}\,$ M$_{\odot}$. In this case, the ratio of masses is given
by $m_p/m_r=$0.5 and again the desired resolution is achieved in these simulations.
\subsection{The models and their calibration}
\label{subs:mod}

It must be mentioned that the values of the ratio
$\left(\frac{E_{\rm kin}}{E_{\rm grav}}\right)_{\rm max}$, defined in Section \ref{subs:energies}, were determined by using an
iterative process, so that for a given value of $\frac{E_{\rm kin}}{E_{\rm grav}}$, it was verified that the model
collapses, otherwise a new  value of $\frac{E_{\rm kin}}{E_{\rm grav}}$, higher than the previous one, was used until the model
no longer collapses.

A clarification is needed to understand this calibration process of the models.
As was mentioned in Section~\ref{subsec:velfin}, in order to determine the
particle velocity, a Fourier mesh of $N_g$ elements per side was used to generate
the turbulent modes; this implies the calculation of $3 \times N_p \, \times \, N_g^3$ operations
where $N_p$ is the total number of particles, see Section \ref{subs:resol}. Therefore, at least
$6.3 \times 10^{13}$ operations must be performed to determine the velocity field of a simulation.

Our initial conditions code was parallelized in the number of
particles, so that a processor takes only a fraction of the total
$N_p$ particles to compute their velocities. Running in 80
processors in the cluster Intel Xeon E5-2680 v3 at 2.5 Ghz of
LNS-BUAP, this code takes up to 8 hours to generate the initial
conditions of all the simulation particles.

In such a situation, it is almost impractical to make
the calibration of the models using all the simulation particles.
For this reason, test models of a very few thousand particles were
used, but only to determine the $\frac{E_{\rm ther}}{E_{\rm grav}}$ 
and $\left(\frac{E_{\rm kin}}{E_{\rm grav}}\right)_{\rm max}$ needed to
calibrate each model.

The important point here is that it was verified that the values
of $\frac{E_{\rm ther}}{E_{\rm grav}}$ and $\frac{E_{\rm kin}}{E_{\rm grav}}$ obtained in this test 
calibration do not change significantly in the
complete simulation of $N_p$ particles. However, there is an indetermination factor $\delta$
in $\left(\frac{E_{\rm ther}}{E_{\rm grav}}\right)_{\rm max}$, so that it is expected that the turbulent core
of the complete simulation is still in the collapsing regime within the
range $\left( \frac{E_{\rm kin}}{E_{\rm grav}}\right)_{\rm max} \pm \delta $. In this paper, the upper bound of $\delta$
is less than (or equal to) 0.1.

The models considered in this paper are summarized in
Table~\ref{tab:mod}, whose entries are as follows. Column 1 shows
the model number;  column 2 shows the value of the constant
$C_R$ as defined in Eq.~\ref{defCR}, which determines the size of the
Fourier mesh; column 3 shows the maximal energy ratio $\left( \frac{E_{\rm kin}}{E_{\rm grav}}\right)_{\rm max}$ reached
by the model; column 4 shows the average Mach number\footnote{Defined as the ratio 
of the velocity magnitude to the
sound speed, $v/c_0$.}${\cal M}$ obtained from
the initial snapshot; column 5 shows the speed of sound used
for each model; column 6 shows the number of the figure of the resulting
configuration, and finally, column 7 shows a comment about the type
of configuration obtained.

\section{Results}
\label{sec:results}

The results of each model are illustrated by
a column density plot, in which all the particles are used in order to
make a 3D rendered image taken at the last snapshot available for
each model. A bar located at the bottom of each iso-density plot
shows the range of values for the $log$ of the column density
$\rho(t)$, calculated in code units by the program {\it splash}
version $2.7.0$, see \cite{splash}. The density unit is given by uden=4.77 $\times
10^{-21}$, so that the average density in code units is
$\rho_0$/uden = 1.29. The colour bar shows values typically in the
range 0--9, so that the peak column density is 10$^9 \, \times $ uden
= 4.77 $ \times 10^{-12}$ g cm$^{-3}$.

It must  be clarified that the vertical and horizontal axes of all
the iso-density plots indicate the length in terms of the radius
$R_0$ of the core (approximately 3335 AU). So, the Cartesian axes
$X$ and $Y$ vary initially from -1 to 1. However, in order to
facilitate the visualization of the last configuration obtained, we
do not use the same length scale per side in all the plots.
\subsection{Dynamical evolution: A general picture}
\label{subsec:modU}

We now briefly describe, in general terms, the time evolution of the
core as seen by using the publicly available visualization code
splash, see \cite{splash}.

\begin{figure}
\begin{center}
\includegraphics[width=4.0 in]{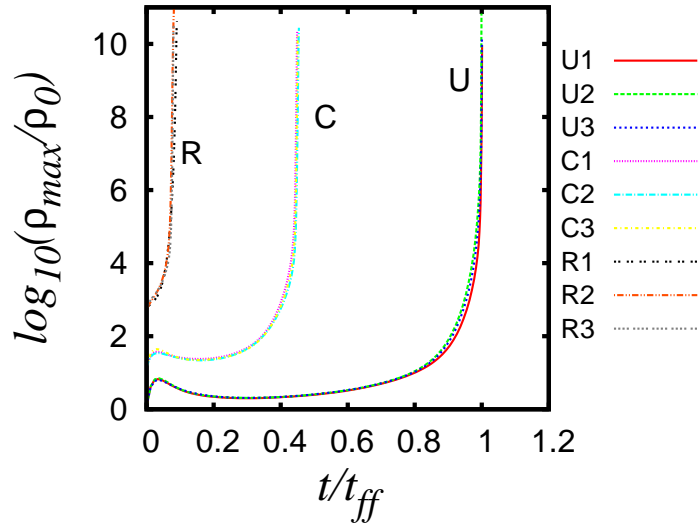}
\caption{\label{fDenMax} Time evolution of the peak density $\rho_{max}$ of the models U, C and R. The curves overlap.}
\end{center}
\end{figure}

A turbulent velocity field favours the occurrence of collisions
between particles across the entire core volume.
We decided to include the turbulence only at the initial
simulation time and to leave the core to collapse freely, so that the first
evolution stage is characterized by an early increase in the
peak density, which can be seen in the first part of the curves
for the U and C models shown in Fig.~\ref{fDenMax}. After this density rebound, the core
truly begins its collapse, because the initial kinetic energy of the turbulence
velocity field was already dissipated.

\begin{figure}
\begin{center}
\includegraphics[width=5 in]{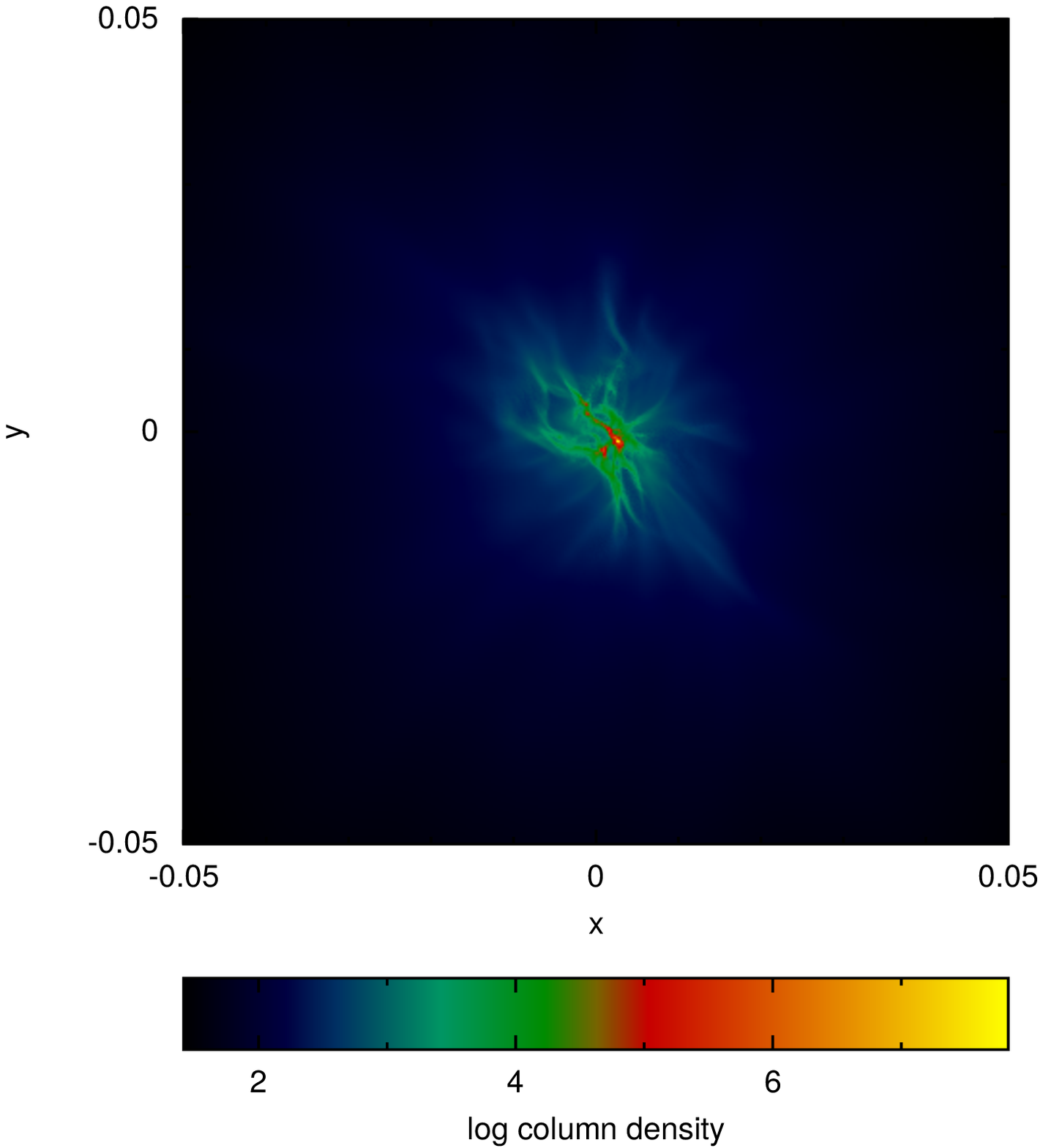}
\caption{\label{fMod1} Column-density plot of the central $(-0.05/R_0, 0.05/R_0)$ core X-Y midplane for model U1
when $\rho_{\rm max}=5.7 \times 10^{-11}$ g/cm$^{3}$ and $t/t_{ff}=1.0$.}
\end{center}
\end{figure}

\begin{figure}
\begin{center}
\includegraphics[width=5 in]{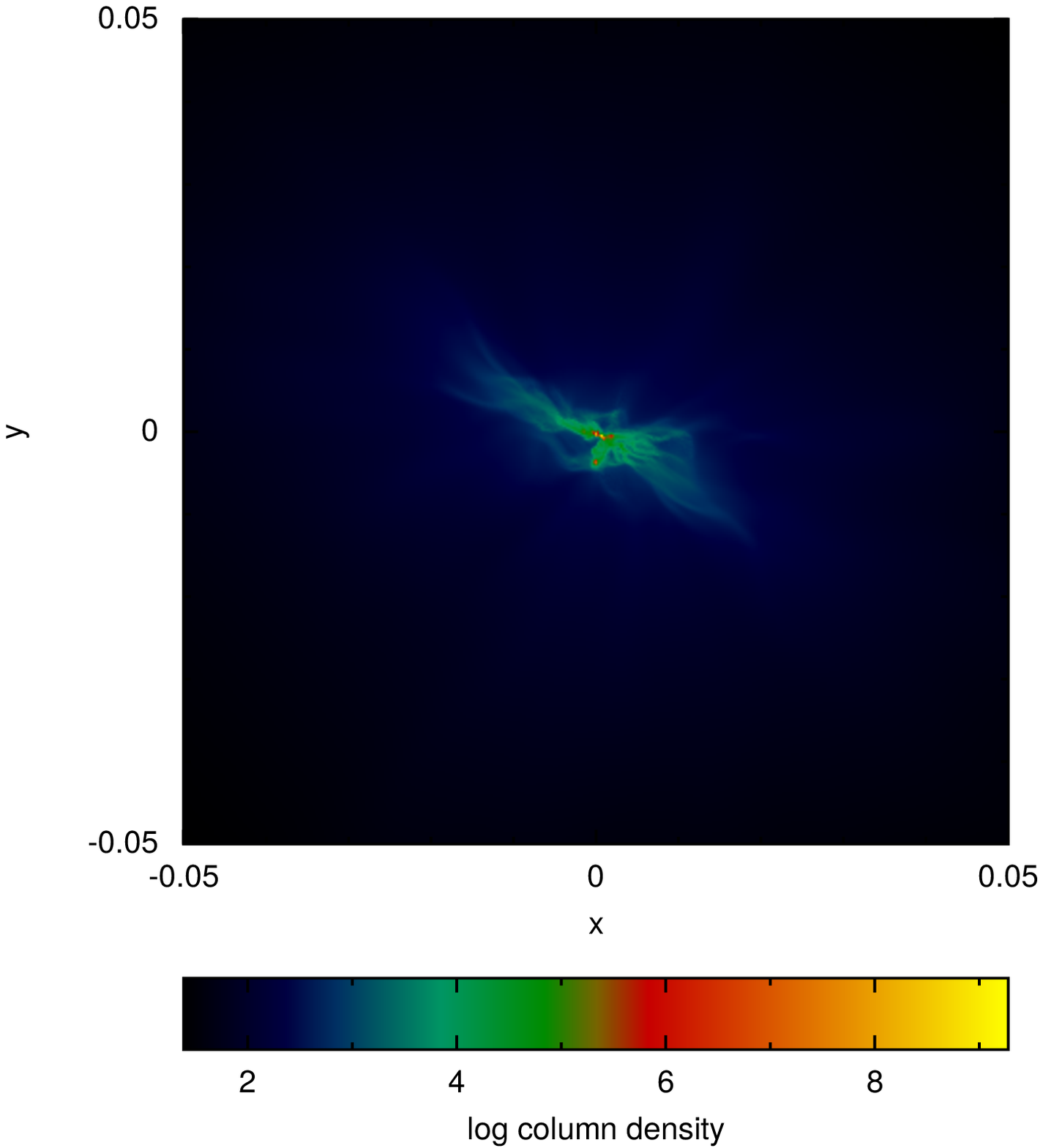}
\caption{\label{fMod4} Column-density plot of the central $(-0.05/R_0, 0.05/R_0)$ core X-Y midplane for model U2
when $\rho_{\rm max}= 4.1 \times 10^{-10}$ g/cm$^{3}$ and $t/t_{ff}=0.99$.}
\end{center}
\end{figure}

\begin{figure}
\begin{center}
\includegraphics[width=5 in]{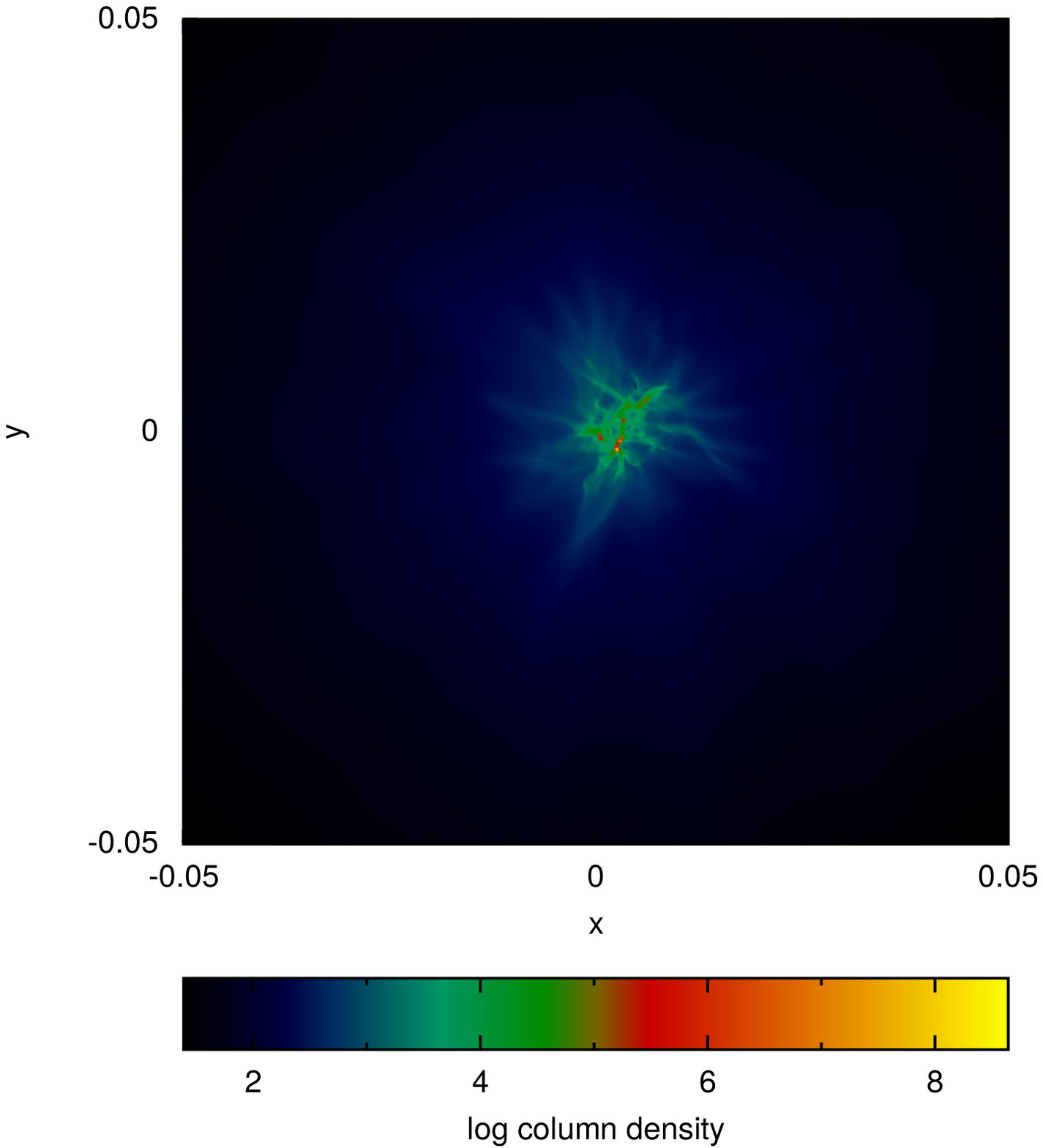}
\caption{\label{fMod10} Column-density plot of the central $(-0.05/R_0, 0.05/R_0)$ core X-Y midplane for model U3 when
$\rho_{\rm max}= 9.7 \times 10^{-11}$ g/cm$^{3}$ and $t/t_{ff}=1.0$.}
\end{center}
\end{figure}

For the uniform models, it can be seen that many dense gas filaments are formed
at the central region of the core. Some of them show fragmentation, so that a few
protostars are in their way of formation, see Fig.~\ref{fMod1}. When the wavelength of the perturbation
mode increases to $4 \times R_0$, it can be seen that one direction is dominant, along which the filaments
form, see Fig.~\ref{fMod4}. In the case of model U3, the central overdensity seems to be a little
bit elongated, see Fig.~\ref{fMod10}.

The density rebound, characterizing the first stage of evolution is not visible in the
peak density curves of models R. The
reason for this is that, at time
$t=0$, all the particles are uniformly distributed across the entire
core volume; for the uniform models, all the particles have
the same mass, whereas for the centrally condensed models, those
particles located in the innermost central region have larger masses
than those located in the outermost core. The collapse is thus
stronger and quicker in the central region of the centrally condensed core.

\begin{figure}
\begin{center}
\includegraphics[width=5 in]{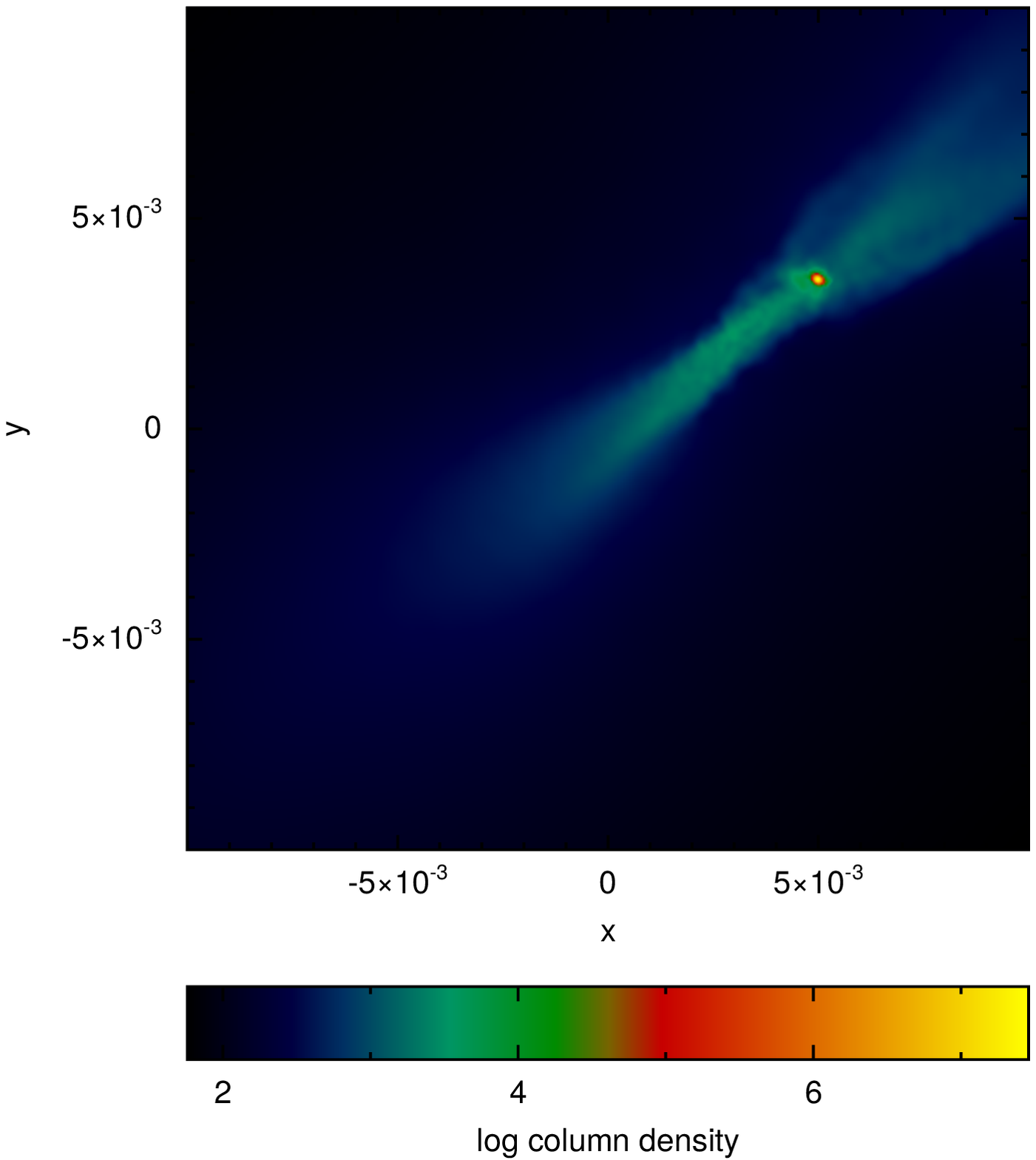}
\caption{\label{fRCen1} Column-density plot of the central $(-0.01/R_0, 0.01/R_0)$ core X-Y midplane for model C1 when
$\rho_{\rm max}= 1.8 \times 10^{-10}$ g/cm$^{3}$ and $t/t_{ff}=0.449$.}
\end{center}
\end{figure}

\begin{figure}
\begin{center}
\includegraphics[width=5 in]{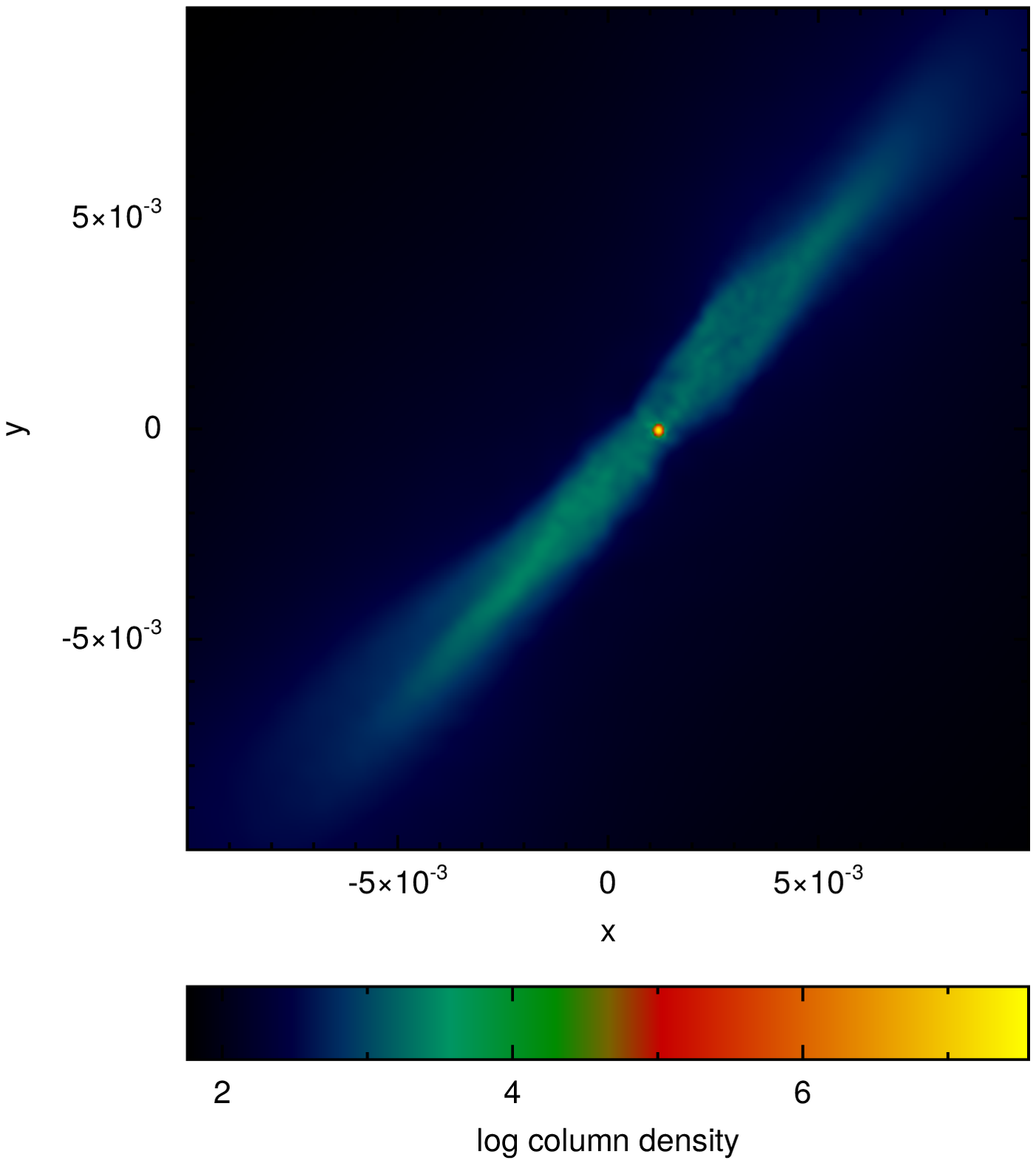}
\caption{\label{fRCen4} Column-density plot of the central $(-0.01/R_0, 0.01/R_0)$ core X-Y midplane for model C2 when
$\rho_{\rm max}= 1.69 \times 10^{-10}$ g/cm$^{3}$ and $t/t_{ff}=0.454$.}
\end{center}
\end{figure}

\begin{figure}
\begin{center}
\includegraphics[width=5 in]{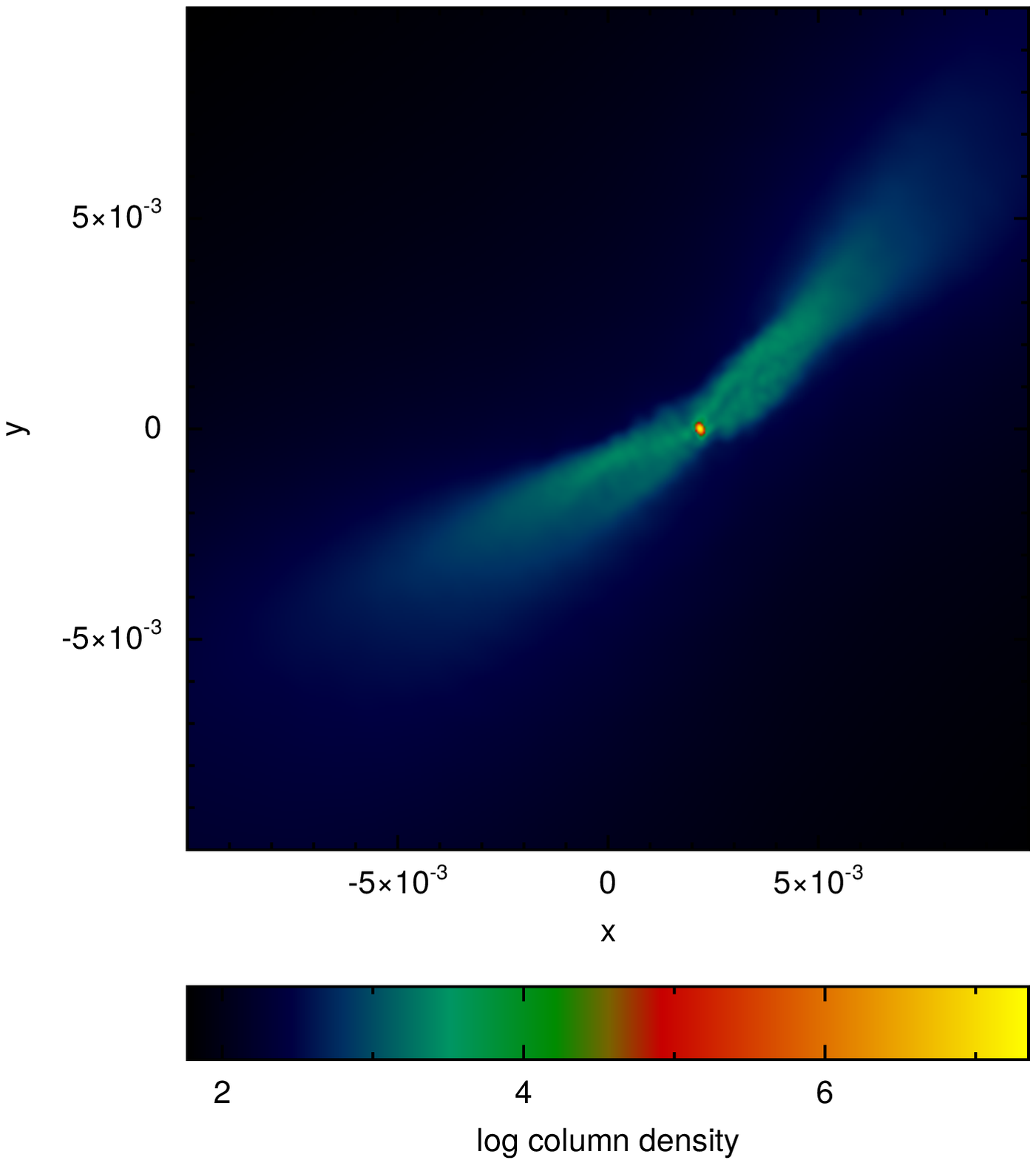}
\caption{\label{fRCen10} Column-density plot of the central $(-0.01/R_0, 0.01/R_0)$ core X-Y midplane for model C3 when
$\rho_{\rm max}= 1.04 \times 10^{-10}$ g/cm$^{3}$ and $t/t_{ff}=0.45$.}
\end{center}
\end{figure}

For the C models, we observe mainly the formation of a long filament, inside
which there is a well-defined overdensity in the central region of the core, so in these models, the
collapse of the core will form only a primary protostar inside a gas filament, see
Fig.~\ref{fRCen1}, Fig.~\ref{fRCen4} and Fig.~\ref{fRCen10}.

\begin{figure}
\begin{center}
\includegraphics[width=5 in]{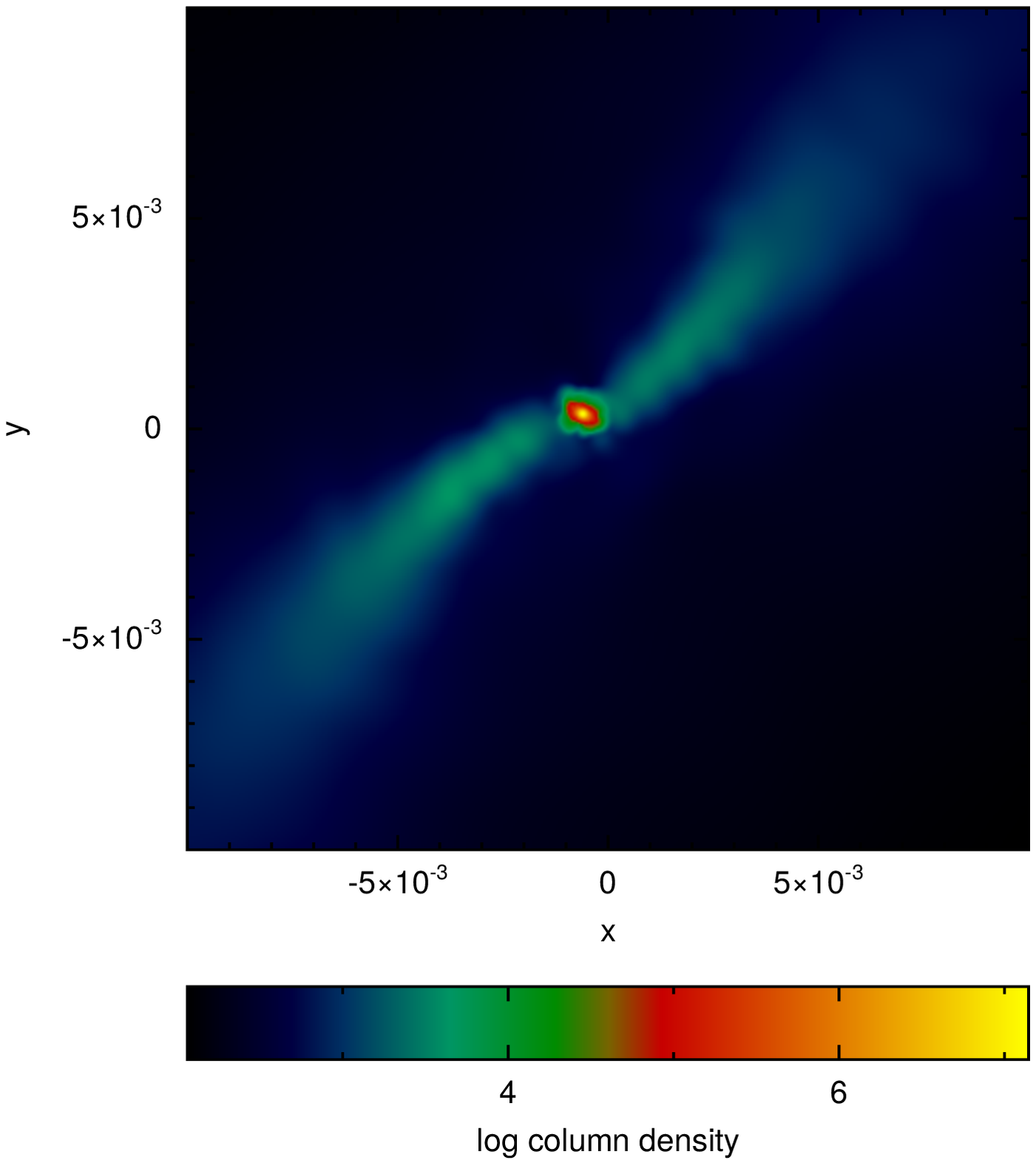}
\caption{\label{fCen1} Column-density plot of the central $(-0.01/R_0, 0.01/R_0)$ core X-Y midplane for model R1 when
$\rho_{\rm max}=2.5 \times 10^{-10}$ g/cm$^{3}$ and $t/t_{ff}=0.09$.}
\end{center}
\end{figure}

\begin{figure}
\begin{center}
\includegraphics[width=5 in]{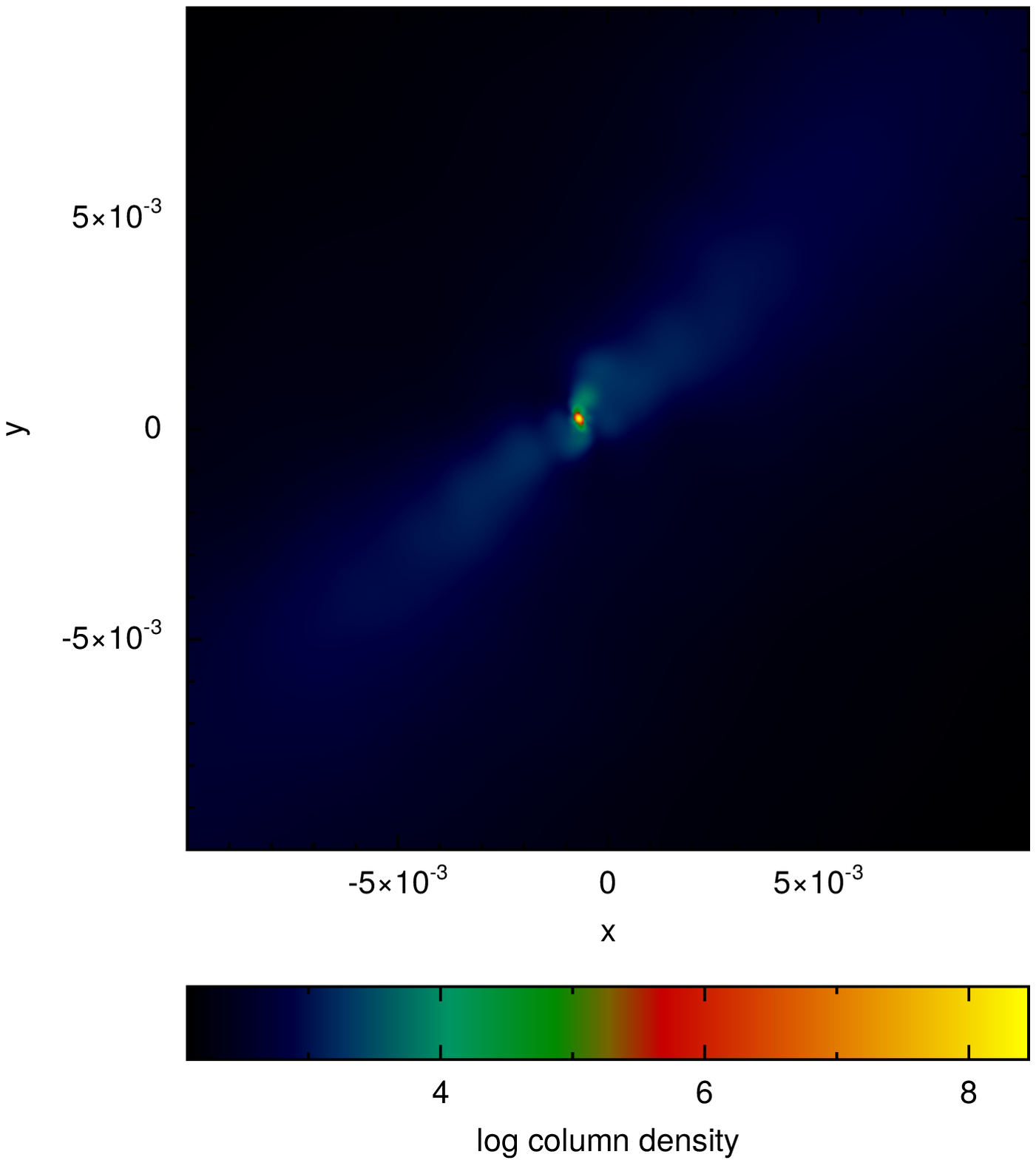}
\caption{\label{fCen4} Column density plot of the central $(-0.01/R_0, 0.01/R_0)$ core X-Y midplane for model R2 when
$\rho_{\rm max}= 1.9 \times 10^{-9}$ g/cm$^{3}$ and $t/t_{ff}=0.08$.}
\end{center}
\end{figure}

\begin{figure}
\begin{center}
\includegraphics[width=5 in]{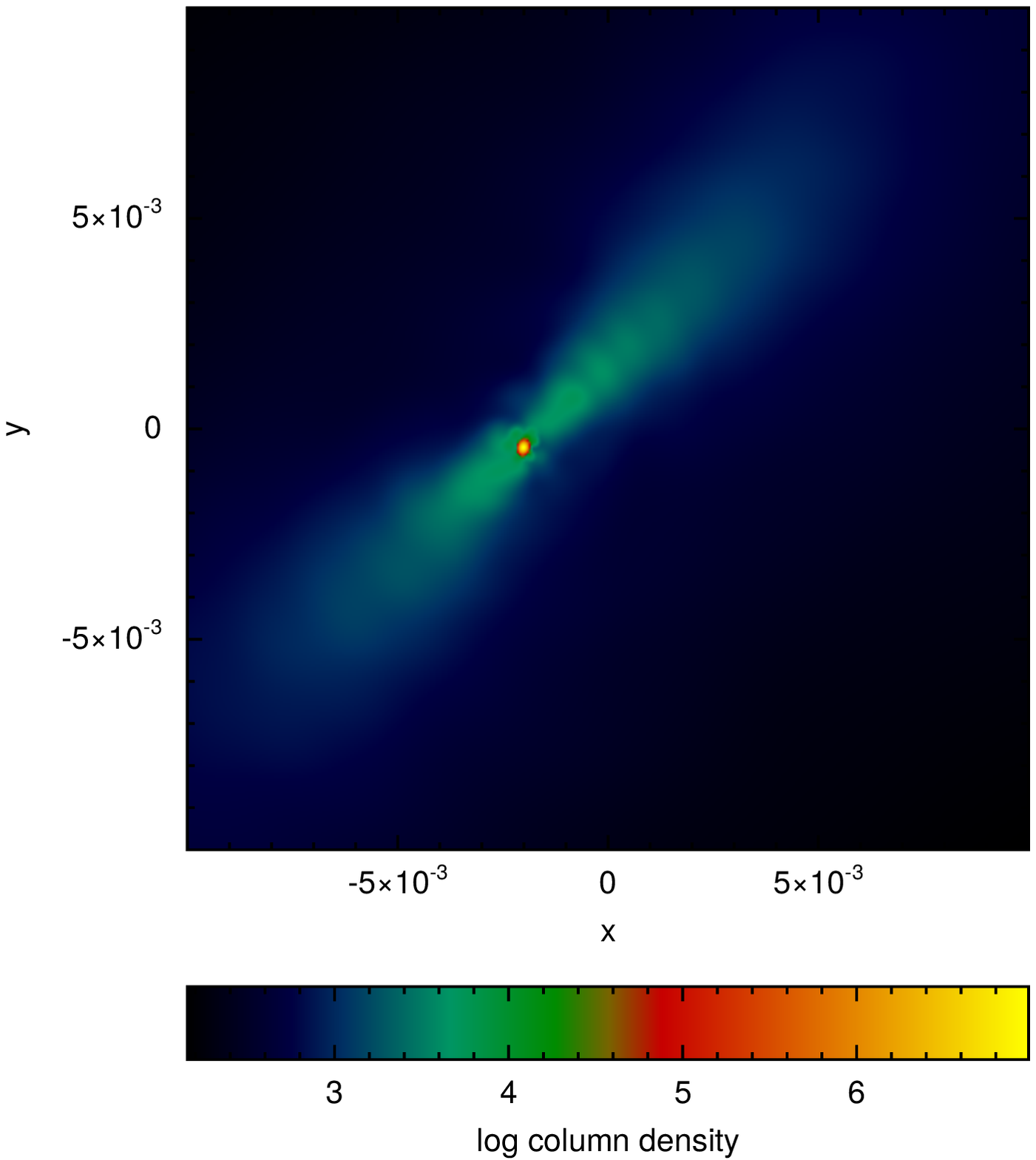}
\caption{\label{fCen10} Column-density plot of the central $(-0.01/R_0, 0.01/R_0)$ core X-Y midplane for model R3 when
$\rho_{\rm max}= 2.0 \times 10^{-10}$ g/cm$^{3}$ and $t/t_{ff}=0.08$.}
\end{center}
\end{figure}

In the case of R models, the central region of the core accretes much more
mass very rapidly and thus the collapse takes place more rapidly in
the central region, so that the mentioned filament could have a net angular momentum, see
and compare Fig.~\ref{fCen1}, Fig.~\ref{fCen4} and Fig.~\ref{fCen10}.

For all the centrally condensed models, the central collapse is so strong that the increase of the
wavelength of the perturbation mode does not affect significantly the basic configuration.
\subsection{Physical properties}
\label{subsec:intprop}

In order to calculate some physical properties of the resulting
protostars, such as the mass and the values of the energy ratios
$\left( \frac{E_{\rm ther}}{E_{\rm grav}}\right)_f$ and $\left(\frac{E_{\rm kin}}{E_{\rm grav}}\right)_f$, we used 
a subset of the simulation
particles, which are determined by means of the following procedure.
First, the highest density particle of the last available snapshot for each
model was located; this particle is going to be the centre
of the protostar. All those particles are found that (i) have density above
some minimum value fixed beforehand $\log_{10} \left( \rho_{\rm
min}/\rho_0 \right)=2.0$ for all the turbulent models and (ii) are also
located within a given maximum radius $r_{\rm max}$ from the
protostar's centre.

The physical properties thus obtained are reported in Table
\ref{tab:propint}, whose entries are as follows. The first column
shows the number and label of the model; the second column shows the
parameter $r_{\rm max}$ given in terms of the core radius $R_0$; the
third column shows the number of particles used in the calculation of
the properties; the fourth column shows the mass of the resulting
protostar given in terms of $M_\odot$; the fifth and sixth columns
give the values of $\left(\frac{E_{\rm ther}}{E_{\rm grav}}\right)_f$ 
and $\left(\frac{E_{\rm kin}}{E_{\rm grav}}\right)_f$, respectively.

Many lines are included in the Table \ref{tab:propint} as protostars have been
found in each uniform model, so that the properties of different protostars are referred to in
different lines of Table \ref{tab:propint}. The average separation of the protostar
found in the uniform models is around 23 to 29 AU (astronomical units).

\begin{figure}
\begin{center}
\includegraphics[width=4.0 in,height=4.0 in]{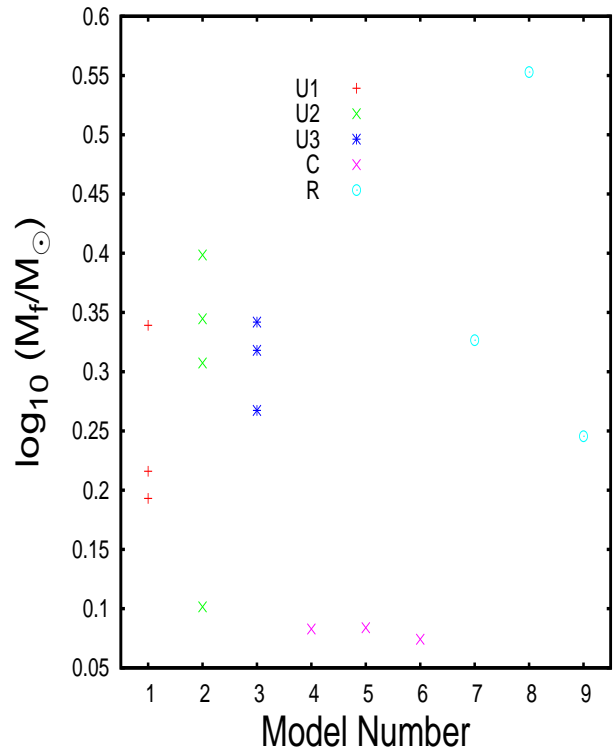}
\caption{\label{MassFrags} Fragment masses against the model number; see Table~\ref{tab:propint}.}
\end{center}
\end{figure}

In Fig. \ref{MassFrags}, the masses of the protostars are shown in
terms of the model number. For the U models, the masses ranges within 0.2-0.4 $M_\odot$. For C models, the
masses are less than 0.1 $M_\odot$. As expected, the largest protostar mass was
found for one of the R models.

\begin{figure}
\begin{center}
\includegraphics[width=4.0 in,height=4.0 in]{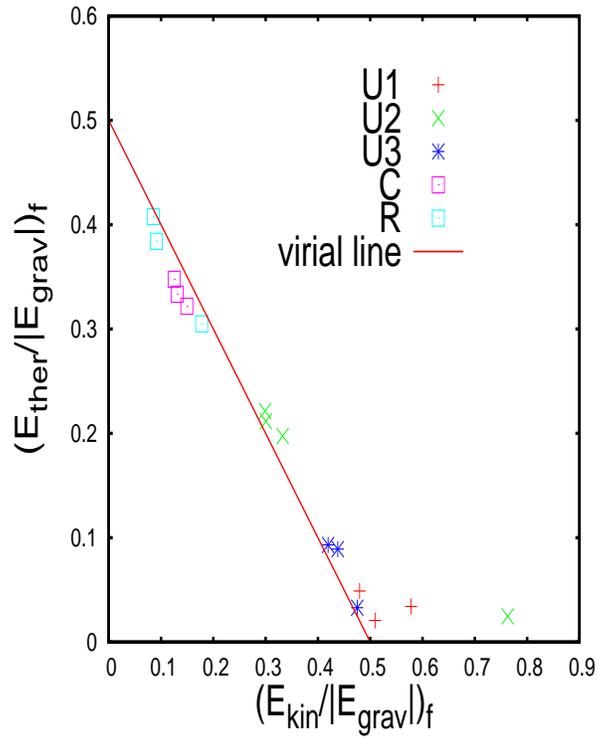}
\caption{\label{AlphavsBetaFrag} Integral properties of protostars: the vertical axis is the ratio of the thermal energy to the gravitational
energy and the horizontal axis is the ratio of the kinetic energy
to the gravitational energy; see columns 5 and 6 of Table~\ref{tab:propint}.}
\end{center}
\end{figure}

It must be mentioned that the values obtained for the energy ratios
of the protostars, namely $\left(\frac{E_{\rm ther}}{E_{\rm grav}}\right)_f$ and 
$\left(\frac{E_{\rm kin}}{E_{\rm grav}}\right)_f$, unfortunately
depend on the values chosen for the parameters $\rho_{\rm min}$ and
$r_{\rm max}$, as there is an ambiguity in defining the boundaries of
the protostar. Despite this, the values of $\left(\frac{E_{\rm ther}}{E_{\rm grav}}\right)_f$ and
$\left( \frac{E_{\rm kin}}{E_{\rm grav}}\right)_f$ for the C models show a clear tendency to virialize, as can
be appreciated in Fig.~\ref{AlphavsBetaFrag}. This behavior is
to be expected, as the collapse of these models was so strongly centrally
dominated that the formed protostars reach very quickly a thermal
support against gravity. We emphasize that a similar observation was made by ~\cite{RMAA2012}.
\section{Discussion}
\label{sec:dis}

In this paper, we have studied the nature of the gravitational collapse of a core
when extreme values of kinetic energy are provided initially by means of several turbulent
velocity fields.

In order to make this paper of interest for star formation,
in spite of the fact that these high values of $\left(\frac{E_{\rm kin}}{E_{\rm grav}}\right)_{\rm max}$ are 
perhaps unrealistic initial
conditions in the observational sense, we always focused on collapsing cores and so
we thus showed that the parameter space of $\frac{E_{\rm kin}}{E_{\rm grav}}$ is enormous. Indeed, in
Section~\ref{subs:mod} we explained in detail
the calibration process of our models in order to obtain the maximal
value of $\left( \frac{E_{\rm kin}}{E_{\rm grav}}\right)_{\rm max}$ allowed in a collapsing core, namely: we always checked the
occurrence of the collapse of the core before we increased the value of $\left(\frac{E_{\rm kin}}{E_{\rm grav}}\right)_{\rm max}$, so when the core did not collapse anymore, the process was
stopped.

According to Fig.~\ref{fMaxBeta}, where we show the $\left(\frac{E_{\rm kin}}{E_{\rm grav}}\right)_{\rm max}$ 
obtained for each model against
the central density of the core ( see also column 3 of Table~\ref{tab:mod}), the uniform models
can manage a very high input of initial energy, which is dissipated by means of particle
collisions, so that the collapse takes place around a 1 free fall
time $t_{ff}$, see Fig.~\ref{fDenMax}. At the opposite extreme, the centrally condensed models need much less input
of initial energy, so that their collapse take place very quickly: for instance the collapse
time for C models ranges around 0.45 times $t_{ff}$, while for R models, the collapse time varies
within 0.08-0.09 times $t_{ff}$.

This behavior is to be expected, as the local free-fall time of C and R models
in the innermost region of the core are shorter than in the
outermost region. In these cases, we observed the occurrence of the so-called inside-out gas
collapse, in which the central region of the core collapses first while
much gas is left behind with low levels of collapse in the core
outer region.

It should be noted that the values thus obtained for
$\left(\frac{E_{\rm kin}}{E_{\rm grav}}\right)_{\rm max}$ (recall that $\frac{E_{\rm ther}}{E_{\rm grav}}$ was fixed for all the models ) were so high
that the collapse of the core would not be expected, because the sum
of the values chosen for the initial energy ratios $\frac{E_{\rm ther}}{E_{\rm grav}}$ and
$\left( \frac{E_{\rm kin}}{E_{\rm grav}}\right)_{\rm max}$ is always higher than the equilibrium value given
in Eq.~\ref{abvirial}, and therefore these ratios would not favour in
principle the global collapse of the core, see Section
\ref{subs:energies}. Two clarifying comments are in order concerning the
high values of $\left(\frac{E_{\rm kin}}{E_{\rm grav}}\right)_{\rm max}$.

Firstly, it must be taken into account that the virial theorem of
Section \ref{subs:energies} does not include the effects of
turbulence on the dynamics of the core, namely, that many gas
collisions take place simultaneously across the core in the initial
stage of the simulation, so that the gas is simultaneously
compressed in many places, favouring the local collapse of the core.
In fact, early theoretical attempts to take into account this
particular nature of turbulence in the task of predicting the fate
of a turbulent gas system were made by \cite{sasao} and
\cite{bona}, who suggested considering a wavenumber-dependent
effective speed of sound, so that $c_0 \equiv c_0(K)$. Because of this
issue, it is also possible that the resolution analysis presented in
Section~\ref{subs:resol} is incomplete, in the sense that a Jeans
mass $M_J$ must be replaced as well by an effective Jeans mass $M_J
\equiv M_J(K)$.

Secondly, observational values of $\frac{E_{\rm kin}}{E_{\rm grav}}$ for prestellar cores
have been found to be within the range 10$^{-4}$ to 0.07, see \cite{caselli} and
\cite{jijina}\footnote{The possibility that other observations have values outside
this range is still open.}. It thus seems that the high values of $\frac{E_{\rm kin}}{E_{\rm grav}}$ of this paper
are not physically justified. This is a very important point, but it must be
mentioned that the initial conditions for the standard isothermal test
case calculated by \cite{boss1979}, \cite{boss1991}, \cite{truelove98},\cite{klein99},
\cite{boss2000}, and \cite{kitsionas}, among others, were $\frac{E_{\rm ther}}{\left|E_{\rm grav}\right|}=0.26$
and $\frac{E_{\rm kin}}{\left|E_{\rm grav}\right|}=0.16$.

A possible limitation of the models presented in this paper is that we did not
use the sink technique introduced by
\cite{batebonnellprice95} and later updated by \cite{fede2010}, so that our simulations did not evolve
much further in time than when the collapse occurred. If we were able to follow the
simulations longer, we could possibly see the fragmentation of the
filament of some primary protostars.

A final consideration to be mentioned here is that the stochastic
nature of the turbulent spectrum of the velocity field was not taken into
account, namely, in this paper, the seed to generate the random numbers was
the same for all simulations and it was fixed from the beginning. However, it has
been shown elsewhere that simulations with different realizations of the random seeds can have
significant differences in their outcomes; see for instance \cite{walch}, \cite{giri2011}, \cite{fede2012}, who used
different random seeds to make a suite of turbulent simulations.

\section{Concluding Remarks}
\label{sec:conclu}

In this paper we carefully prepared the initial conditions for the particles in order to have a collapsing core
and thus we bounded the value of $\left( \frac{E_{\rm kin} }{E_{\rm grav} }\right)_{\rm max}$ for each model, so that beyond this
value of $\frac{E_{\rm kin}}{E_{\rm grav}}$, the core get dispersed unboundedly. We emphasized that the total mass and
radius of the core are the same
for all the models, so that we changed only (i) the radial density profile and (ii) the wavelength of the
perturbation mode.

\begin{figure}
\begin{center}
\includegraphics[width=4.0 in,height=4.0 in]{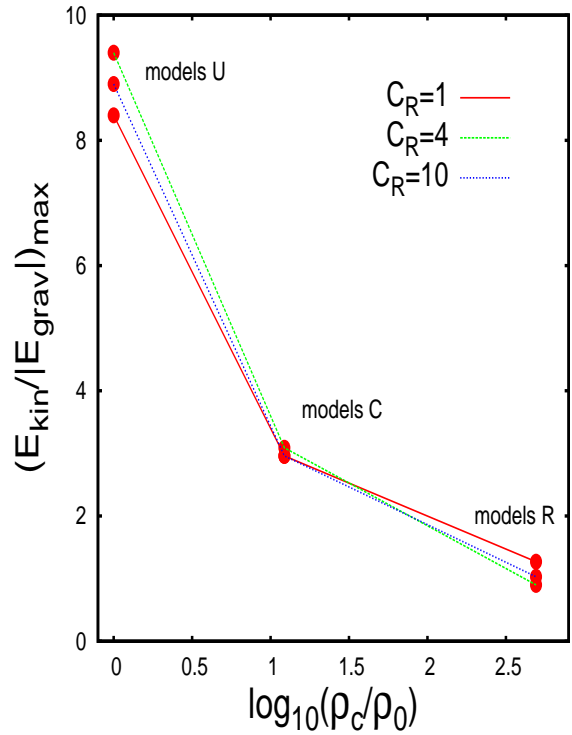}
\caption{\label{fMaxBeta} The value of $\left(\frac{E_{\rm kin}}{E_{\rm grav}}\right)_{\rm max}$ against the central density of the core (normalized to the average density of the core $\rho_0$). The lines join models with the same wavelength of the perturbation mode and the points show
the measurements made in this paper.}
\end{center}
\end{figure}

The main result of this paper is therefore shown in Fig.\ref{fMaxBeta}, in which we observed that (i)
less input of kinetic energy is required to disperse the core as the central density increases; (ii)
the effect of the increase in the wavelength of the perturbation mode is very small.

We obtained  interesting configurations at the end of the core
collapse, particularly those of models 1-3, which showed multiple fragmentation mainly along 
filaments. \citet{lomax} studied a total of 50 models of velocity fields for core collapse in which the
turbulent energy was varied, and identified two
principal modes of fragmentation, namely the filament fragmentation and
disk fragmentation. It must be noted that configurations of models 1-3 of this paper, correspond
to the filament fragmentation mode.

In addition, the centrally condensed models resulted in the formation of  a single central 
protostar, which is inside a long filament. Fragmentation was perhaps suppressed by the extreme 
mass concentration at the centre of the core. 

\acknowledgments

The author thankfully acknowledge the computer resources, technical expertise and support provided by the
Laboratorio Nacional de Superc\'omputo del Sureste de M\' exico through the grant number O-2016/047.
\newpage

\clearpage

\begin{table}[ph]
\caption{Turbulent models and their resulting configurations}
{ \begin{tabular}{|c|c|c|c|c|c|c|c|c|}
\hline \hline
Model Number (label)  & $C_R$ & $\left(\frac{E_{\rm kin}}{\left|E_{\rm grav}\right|}\right)_{\rm max}$  & ${\cal M}_{\rm ave}$    & $c_0$ [cm/s] & Figure        & Configuration         \\
1 (U1)     & 1     &  8.4   &   8.96                       & 9677.64  & \ref{fMod1}   & primary + fragmented filaments \\
2 (U2)     & 4     &  9.4   &   9.45                       & 9677.64  & \ref{fMod4}   & fragmented filaments          \\
3 (U3)     & 10    &  8.9   &   9.24                       & 9677.64  & \ref{fMod10}  & fragmented filaments          \\
\hline
4 (C1)     & 1     &  2.96  &   5.38                       & 11423.37 & \ref{fRCen1}  & single + long filament    \\
5 (C2)     & 4     &  3.09  &   5.49                       & 11423.37  & \ref{fRCen4} & single + long filament        \\
6 (C3)     & 10    &  2.96  &   5.37                       & 11423.37 & \ref{fRCen10} & single + long filament      \\
\hline
7 (R1)     & 1     &  1.27  &   3.63                       &  17097.9 & \ref{fCen1}   &  single + long filament       \\
8 (R2)     & 4     &  0.9   &   3.05                       &  17097.9 & \ref{fCen4}   &  single + long filament \\
9 (R3)     & 10    &  1.03  &   3.2                        &  17097.9 & \ref{fCen10}  &  single + long filament    \\
\hline
\end{tabular} }
\label{tab:mod}
\end{table}

\begin{table}[ph]
\caption{Integral properties of protostars}
{\begin{tabular}{|c|c|c|c|c|c|}
\hline\hline
Model Number (label) & r$_{\rm max}/R_0$    &   $N_p$     & $M_f/M_\odot$ & $\left(\frac{E_{\rm ther}}{\left|E_{\rm grav}\right|}\right)_{f}$  & $\left(\frac{E_{\rm kin}}{\left|E_{\rm grav}\right|}\right)_{f} $  \\
1 (U1)    &   0.008           & 628984  & 3.39e-01   &    0.049  &  0.479 \\
1 (U1)    &   0.008           & 400546  & 2.16e-01   &    0.020  &  0.509 \\
1 (U1)    &   0.008           & 358062  & 1.93e-01   &    0.034  &  0.578 \\
\hline
2 (U2)    &   0.008           & 739098  &  3.98e-01  &    0.197  &  0.333 \\
2 (U2)    &   0.008           & 639297  &  3.45e-01  &    0.211  &  0.300 \\
2 (U2)    &   0.008           & 570112  &  3.07e-01  &    0.221  &  0.299 \\
2 (U2)    &   0.008           & 188238  &  1.01e-01  &    0.025  &  0.762 \\
\hline
3 (U3)    &   0.008            & 634095  &  3.42e-01  &    0.089  &  0.438 \\
3 (U3)    &   0.008            & 589909  &  3.18e-01  &    0.093  &  0.419 \\
3 (U3)    &   0.008            & 495728  &  2.67e-01  &    0.033  &  0.475 \\
\hline
4 (C1)    &   0.005            & 15187   &  8.28e-02  &    0.322  &  0.149 \\
5 (C2)    &   0.005            & 15419   &  8.39e-02  &    0.348  &  0.125 \\
6 (C3)    &   0.005            & 13450   &  7.41e-02  &    0.333  &  0.130 \\
\hline
7 (R1)    &   0.005            & 2514    &  3.27e-01  &    0.305  &  0.178  \\
8 (R2)    &   0.005            & 4310    &  5.52e-01  &    0.408  &  0.085 \\
9 (R3)    &   0.005            & 1666    &  2.46e-01  &    0.384  &  0.090 \\
\hline
\end{tabular}}
\label{tab:propint}
\end{table}
\newpage
\newpage

\begin{thebibliography}{}

\bibitem[Andr\'e (2007)]{andre} Andr\'e P., Belloche, A., Motte, F. and Peretto, N., 2007, Astron. Astrophys., 472, 519.

\bibitem[Arreaga et al. (2009) ]{miAA} Arreaga, G., Klapp, J. and Gomez, F., 2009, Astron. Astrophys., 509, pp. A96--A111.

\bibitem[Arreaga et al. (2007)]{miApJ} Arreaga, G., Klapp, J.,Sigalotti L.D., Gabbasov, R., 2007, \apj, 666, 290--.

\bibitem[Arreaga and Saucedo (2012)]{RMAA2012} Arreaga-Garc\'{\i}a, G. and Saucedo-Morales, J.C., 2012, Revista Mexicana
de Astronom\'{\i}a y Astrof\'{\i}sica, 48, pp. 61--84.

\bibitem[Arreaga (2016)]{RMAA2016} Arreaga-Garc\'{\i}a, G., 2016, Revista Mexicana de Astronom\'{\i}a y
Astrof\'{\i}sica, 52, pp. 155--169.

\bibitem[Arreaga (2017)]{miAAS} Arreaga-Garc\'{\i}a, G., 2017, Astrophysics and Space Science {\bf 362}, pp. 47--.

\bibitem[Arreaga (2017)]{RMAA2017} Arreaga-Garc\'{\i}a, G., 2017, Revista Mexicana de Astronom\'{\i}a y
Astrof\'{\i}sica, 53, pp. 1--24.

\bibitem[Attwood et al. (2009)]{attwood} Attwood, R.E.,  Goodwin, S.P., Stamatellos, D. and Whitworth, A.P.,
2009, Astron. Astrophys., 495, pp. 201--215.

\bibitem[Ballesteros et al. (1999)]{ballesteros1999} Ballesteros-Paredes, J., Hartmann, L. and
Vazquez-Semanedi, E.,  1999, \apj, 527, pp. 285.

\bibitem[Balsara (1995)]{balsara1995} Balsara, D., 1995, J. Comput. Phys., 121, 357.

\bibitem[Bate et al. (1995)]{batebonnellprice95} Bate, M.R., Bonnell, I.A. and Price,
N.M., 1995, \mnras, 277, pp. 362--376.

\bibitem[Bate et al. (2002a)]{bate2002a} Bate, M.R., Bonnell, I.A. and Bromm, V., 2002, \mnras, 332, pp. L65--L68.

\bibitem[Bate et al. (2002b)]{bate2002b} Bate, M.R., Bonnell, I.A. and Bromm, V., 2002, \mnras, 336, pp. 705--713.

\bibitem[Bate et al. (2003)]{bate2003} Bate, M.R., Bonnell, I.A. and Bromm, V., 2003, \mnras, 339, pp. 577--599.

\bibitem[Bate and Burkert (1997)]{bateburkert97} Bate, M.R. and Burkert, A., 1997, \mnras, 288, 1060.

\bibitem[Bergin et al. (2007)]{bergin} Bergin, E. and Tafalla, M., 2007,
Annu. Rev. Astro. Astrophys., 45, 339.

\bibitem[Bodenheimer (1995)]{boden95} Bodenheimer, P., 1995, \araa, 33, 199.

\bibitem[Bonazzola (1987)]{bona} Bonazzola, S., Falgorone, E., Heyvaerts, J., P\'erault, M., and Puget, J.L.,
1987, Astron. Astrophys., 172, 293.

\bibitem[Boss and Bodenheimer (1979)]{boss1979} Boss, A.P. and Bodenheimer, P. 1979, \apj, 234, 289.

\bibitem[Boss (1980)]{boss80} Boss, A.P., 1980, \apj, 242, pp. 699--709.

\bibitem[Boss (1981)]{boss81} Boss, A.P., 1981, \apj, 250, 636644.

\bibitem[Boss (1991)]{boss1991} Boss, A.P., 1991, Nature, 351, pp. 298.

\bibitem[Brunt et al. (2014)]{bruntyfede} Brunt, C.M. and Federrath, C., 2014, \mnras, 442, 1451.

\bibitem[Brunt et al. (2009)]{brunt} Brunt, C.M., Heyer, M.H. and MacLow, M.M., 2009, Astron. Astrophys., 504, 883.

\bibitem[Bouwman (2004)]{bou} Bouwman, A.P., 2004, Astrophysics and Space Science, 292, 325.

\bibitem[Boss et al. (2000)]{boss2000} Boss, A.P., Fisher, R.T., Klein, R. and McKee, C.F., 2000, \apj, 528, 325.

\bibitem[Burkert and Bodenheimer (1993)]{burkertboden93} Burkert, A. and Bodenheimer, P., 1993, \mnras, 264, 798.

\bibitem[Caselli et al. (2002)]{caselli} Caselli, P., Benson, P.J., Myers, P.C.  and Tafalla, M., 2002, \apj, 572, 238.

\bibitem[Chandrasekhar (1951)]{chandra} Chandrasekhar, S., 1951, Proc. R. Soc. London, 210, 26.

\bibitem[Dobbs et al. (2005)]{dobbs} Dobbs, C.L., Bonnell, I.A. and Clark, P.C., 2005, \mnras, 360, pp. 2--8.

\bibitem[Elmegreen (1993)]{elmegreen} Elmegreen, B.G., 1993, in: \textit{Protostars and Planets III}, pp. 97.

\bibitem[Federrath (2015)]{fede2015} Federrath, C.,2015, MNRAS,  450, Issue 4, p.4035-4042.

\bibitem[Federrath et al. (2010)]{fede} Federrath, C., Roman-Duval, J., Klessen, R.S., Schmidt, W.
and Mac Low, M.M., 2010, Astron. Astrophys., 512, A81.

\bibitem[Federrath et al. (2010)]{fede2010} Federrath, C., Banerjee, R. Clark, P. C. and Klessen, R. S., 2010, \apj, 713, Issue 1, 
pp. 269-290.

\bibitem[Federrath and Klessen (2012)]{fede2012} Federrath, C. and Klessen, R.S., 2012, ApJ,761, Issue 2, article id. 156.

\bibitem[Girichidis et al. (2011)]{giri2011} Girichidis, P., Federrath, C., Banerjee, R. and 
Klessen, R. S., 2011, MNRAS, 413, Issue 4, pp. 2741-2759.

\bibitem[Goodman et al. (1993)]{goodman93} Goodman, A.A., Benson, P.J., Fuller,
G.A., Myers, P.C., 1993, \apj, 406, pp. 528--547.

\bibitem[Goodwin et al. (2004a)]{goodwin2004a} Goodwin, S.P., Whitworth, A.P. and
Ward-Thompson, D., 2004a, Astron. Astrophys., 414, pp. 633--650.

\bibitem[Goodwin et al. (2004b)]{goodwin2004b} Goodwin, S.P., Whitworth, A.P. and
Ward-Thompson, D., 2004b, Astron. Astrophys., 423, pp. 169--182.

\bibitem[Goodwin et al. (2006)]{goodwin2006} Goodwin, S.P., Whitworth, A.P. and
Ward-Thompson, D., 2006, Astron. Astrophys., 452, pp. 487--492.

\bibitem[Hachisu and Heriguchi (1984)]{hachisu1} Hachisu, I.
and Heriguchi, Y.,  1984, Astron. Astrophys., 140, 259.

\bibitem[Hachisu and Heriguchi (1985)]{hachisu2} Hachisu, I.
and Heriguchi, Y.,  1985, Astron. Astrophys., 143, 355.

\bibitem[Hopkins (2013)]{hopkins} Hopkins, F.,  2013, \mnras, 430, pp. 1653--1693.

\bibitem[Jijina (1999)]{jijina} Jijina, J., Myers, P.C. and Adams, F.C.,  1999, \apj, 125, pp. 161--236.

\bibitem[Klein et al. (1999)]{klein99} Klein, R. I., Fisher, R. T.,
McKee, C. F. and Truelove, J. K. 1999, in
Numerical Astrophysics 1998, ed. S. Miyama, K. Tomisaka and T. Hanawa
Kluwer, Dordrecht, pp. 131--.

\bibitem[Kitsionas \& Whitworth (2002)]{kitsionas} Kitsionas, S., Whitworth, A.P., 2002, \mnras, 330, 129.

\bibitem[L\'eorat (1990)]{leorat} L\'eorat, J., Passot, T. and Pouquet, A., 1990, \mnras, 243, pp. 293--311.

\bibitem[Lomax et al. (2015)]{lomax} Lomax, O., Whitworth, A.P., and Hubber, D.A., 2015, \mnras, 449, pp.662-669.

\bibitem[Myers (2005)]{myers} Myers, P.C., 2005, \apj, 623, 280.

\bibitem[Miyama et al. (1984)]{miyama} Miyama, S.M., Hayashi, C.
and Narita, S., 1984, \apj, 279, 621.

\bibitem[Monaghan and Gingold (1983)]{mona1983} Monaghan, J.J. and
Gingold, R.A., 1983, J. Comput. Phys., 52, 374.

\bibitem[Padoan and Norlund (2002)]{padoan2002} Padoan, P. and Nordlund, A., 2002, \apj, 576, pp. 870--879.

\bibitem[Padoan et al. (2014)]{padoan2014} Padoan, P., Federrath, C., Chabrier, G., Evans II, N.J., Johnstone,
D., J$\oslash$rgensen, J.K., McKee, C.F. and Nordlund, E.,  2014, Protostars and Planets VI, H. Beuther,
R. S. Klessen, C. P. Dullemond, and T. Henning (eds.), University of Arizona Press,
Tucson, AZ, pp. 77--100. arXiv. 1312.5365.

\bibitem[Plummer (1911)]{plu} Plummer, H.C., 1911, \mnras, 71, 460.

\bibitem[Sigalotti and Klapp (2001)]{jaime} Sigalotti, L. and Klapp,  J., 2001, Int. J. Mod. Phys. D, 10, 115.

\bibitem[Sasao (1973)]{sasao} Sasao, T., 1973, PASJ, 25, 1.

\bibitem[Springel (2005)]{gadget2} Springel, V., 2005, \mnras, 364, 1105.

\bibitem[Springel et al. (2001)]{serial} Springel, V., Yoshida, N. and  White, S.D.M., 2001, New Astronomy, 6, 79.

\bibitem[Price (2007)]{splash} Price, D., 2007, PASA 24(3), pp. 159--173. The code SPLASH was taken
from http://users.monash.edu.au/~dprice/splash/, SPLASH: a free and open source visualisation tool for Smoothed Particle
Hydrodynamics (SPH) simulations, 2015.

\bibitem[Tafalla et al. (1998)]{tafalla1} Tafalla, M., Mardones, D. and Myers, P.C., 1998, \apj, 504, pp. 900--914.

\bibitem[Tafalla et al. (2004)]{tafalla2} M. Tafalla, M., Myers, P.C., Caselli, P. and Walmsley,
C.M., 2004, Astron. Astrophys., 416, pp. 191--212.

\bibitem[Tobin et al. (2012)]{tobin2012} Tobin, J.J, Hartmann, L.
and Chiang, H.F., 2012, Nature, 492, 83.

\bibitem[Tobin at al. (2013)]{tobin2013} Tobin, J.J., Chandler, C., Wilner, D.J.,
Looney, L.W., Loinard, L., Chiang, H.-F., Hartmann, L., Calvet, N., D'Alessio, P.,
Bourke, T.L. and Kwon, W., 2013, \apj {\bf 779}.

\bibitem[Tobin at al. (2016)]{tobin2016} Tobin, J.J.,  Kratter, K.M., Persson, M.V., Looney, L.W., Dunham, M.M.,
Segura-Cox, D., Li, Z.Y., Chandler, C.J., Sadavoy, S.I., Harris, R.J., Melis, C.,  and Pérez, L.M., 2016, Nature, 38, pp. 483--486.

\bibitem[Tohline (2002)]{tohline} Tohline, J.E., 2002, ARA\&A, 40, 349.

\bibitem[Truelove et al. (1997)]{truelove} Truelove, J.K., Klein, R.I.,  McKee, C.F.,
Holliman, J.H.,  Howell, L.H. and  Greenough, J.A., 1997, \apj, 489, L179.

\bibitem[Truelove et~al (1998)]{truelove98} Truelove, J. K., Klein, R. I.,
McKee, C. F., Holliman, J. H., Howell, L. H., Greenough, J. A. and Woods, D. T. 1998, ApJ, 495, 821.

\bibitem[Tsuribe et al. (1999)]{tsuribe1} Tsuribe, T. and Inutsuka, S.-I., 1999,
\apj {\bf 523}, pp. L155--L158. Tsuribe, T. and Inutsuka, S.-I., 1999, \apj, 526, pp. 307--313.

\bibitem[Walch et al. (2012)]{walch} Walch, S., Whitworth, A.P. and  Girichidis, P., 2012, \mnras, 419, pp. 760--770.

\bibitem[Whitehouse and Bate (2006)]{whitehouse}  Whitehouse, S.C. and  Bate, M.R., 2006, \mnras, 367, 32.

\bibitem[Whitworth and Ward-Thompson (2001)]{whithworth} Whitworth, A.P. and Ward-Thompson, D., 2001, \apj, 54, 317.
\end{thebibliography}
\end{document}